\documentclass{article}

\usepackage{epsfig}
\usepackage{newlfont}

\begin{document}

\phantom{x}
\vspace{-2cm}
\hfill TTP99--15

\hfill NIKHEF 99--015

\hfill hep-ph/9906426

\begin{center}
{\Large\bf Parallelizing the Symbolic Manipulation Program FORM
\footnote{talk presented by D.Fliegner at AIHENP'99, Heraklion, Greece,
Apr 12-16 1999.}}
\end{center}
\begin{center}
{\Large D.Fliegner$^\dagger$, A.R\'etey$^\dagger$, 
J.A.M. Vermaseren$^{\ddagger}$}
\end{center}
\noindent
$^\dagger$Institut f\"ur Theoretische Teilchenphysik,
Universit\"at Karlsruhe, D-76128 Karlsruhe, Germany\\
$^{\ddagger}$NIKHEF, P.O. Box 41882, 1009 DB, Amsterdam, The Netherlands
\vspace{3mm}

\begin{center}
{\large\bf Abstract}
\end{center}
After an introduction to the sequential version of FORM and the 
mechanisms behind it we report on the status of our ongoing project 
of its parallelization. An analysis of the parallel platforms used is 
given and the structure of a parallel prototype of FORM is explained.

\vspace{5mm}
\noindent
{\large\bf 1. The Sequential Version of FORM}
\vspace{5mm}

\noindent
FORM \cite{form} is a program for symbolic manipulation of algebraic 
expressions specialized to handle very large expressions of millions of 
terms in an efficient and reliable way. It is used non-interactively by 
executing a program that contains several parts called modules. The 
execution of each module is again divided into three steps:
\begin{itemize}
\item {\bf Compilation:} the input is translated into an internal 
representation.
\item {\bf Generating:} for each term of the input expressions the 
statements of the module are executed. This in general generates a lot 
of terms for each input term.
\item {\bf Sorting:} all the output terms that have been generated are
sorted and equivalent terms are summed up.
\end{itemize}
FORM only allows local operations on single terms, like replacing parts
of a term or multiplying something to it. 
Together with a sophisticated pattern matcher this seemingly strong
limitation allows the formulation of general and efficient algorithms.
The limitation to local operations makes it possible to handle expressions 
as ``streams'' of terms, that can be read in sequentially from a file and 
be worked on one at a time. The generation of terms is done in a way that 
the output terms drop out term by term also. 
These output terms are stored in two intermediate buffers and a temporary 
sortfile in a staged procedure: When the smaller buffer gets full, its 
content is sorted and copied to the larger buffer. Consequently the small
buffer is free to be filled with the next patch of output terms. 
If the larger buffer gets full, its content is again sorted and copied to 
the temporary file, freeing the large buffer.
At the end of the module all of the existing sorted patches residing in the
two buffers and the sortfile have to be merged stage by stage to one single 
sorted output stream of terms which is written to file and used as input 
source of terms for the next module. 
For the sorting of the small buffer a slightly modified merge sort algorithm 
is used; merging sorted patches of terms in the other stages of sorting is 
done with a tree of losers \cite{knuth}.
This results in a tree-like structure for the generation of terms as well 
as for the sorting. 

Of course the first step before parallelizing the program was to profile and
optimize the sequential code. One of the main achievements was a speedup of 
about a factor 2 on the 64-bit architecture of the alpha processors 
by changing the internal used word-length from 16 ( which results in 32-bit 
arithmetic operations) to 32 bits (64-bit arithmetics). Another speedup of 
about 1.5 was simply achieved by experimenting with the compiler 
optimizations. 
Profiling of FORM in typical applications shows that the time needed for 
compiling the program text into the internal representation is not dependent 
on the size of the problem and usually neglectible. Most of the time is 
spent in generating and sorting of terms.

\vspace{5mm}
\noindent
{\large\bf 2. Evaluation of Different Parallel Platforms}
\vspace{5mm}

\noindent
One of the suppositions for the parallelizing of FORM has been not to limit 
the approach by using a too specialized hardware, but instead to use standard 
message passing libraries (MPI(CH) \cite{mpich} and PVM \cite{pvm}) for the 
parallelization that are available on a wide class of architectures and 
should be portable to new and more powerful systems. Usually both message
passing libraries use specialized device drivers underneath to yield maximum 
performance. During the first part of our project the following combinations
of hard-- and software have been used:

\begin{itemize}
\item
\begin{tabular}{l}
DEC alpha workstation cluster running DEC UNIX 4.0D,\\ 
8 processors, 600MHz, 512MB RAM and 2$\times$4GB disk each,\\ 
interconnection: Fast (100 MBit) \& Gigabit (1000 MBit) Ethernet,\\
1.26 GBit Myrinet \cite{myricom}, 
1.26 GBit ParaStation II \cite{parastation2},\\
message passing: MPI(CH) and PVM (over IP and ParaStation II).
\end{tabular}
\item
\begin{tabular}{l}
IBM SP2 running AIX 4.2.1,\\
168 (in total 256) processors, 120MHz, 512MB RAM,\\
interconnection: special low latency switches,\\
message passing: MPI.
\end{tabular}
\item
\begin{tabular}{l}
ALR Quad6 SMP machine running Solaris 2.6,\\
4 processors, 200MHz PentiumPro, 512MB RAM and 2$\times$4GB disks,\\
interconnection: shared memory,\\
message passing: MPICH (over shared memory).
\end{tabular}
\end{itemize}

\noindent
In the case of the DEC alpha cluster different IP drivers and different
implementations of device drivers for the messages libraries exist. For a
thorough understanding of the parallel systems behaviour it is of course 
crucial to compare the performance of the messages passing libraries. 
The MPI(CH) libraries have been examined using the Pallas MPI benchmarks 
(PMB) \cite{pmb}. A shell script is provided that can be used to measure the 
throughput and latency of the message passing operations under different 
circumstances without further interaction. We only present the results for
the simplest possible single transfer benchmark, a ping-pong test:
One process(or) sends a message of $n$ bytes to another process(or), 
which immediately sends that message back. In contrast to more complicated 
benchmarks there is no concurrency with other messages passing activity 
during this test. In other words, the bandwidth and latency are measured 
under optimal conditions.

\begin{figure}[h!]
\begin{center}
\epsfig{file=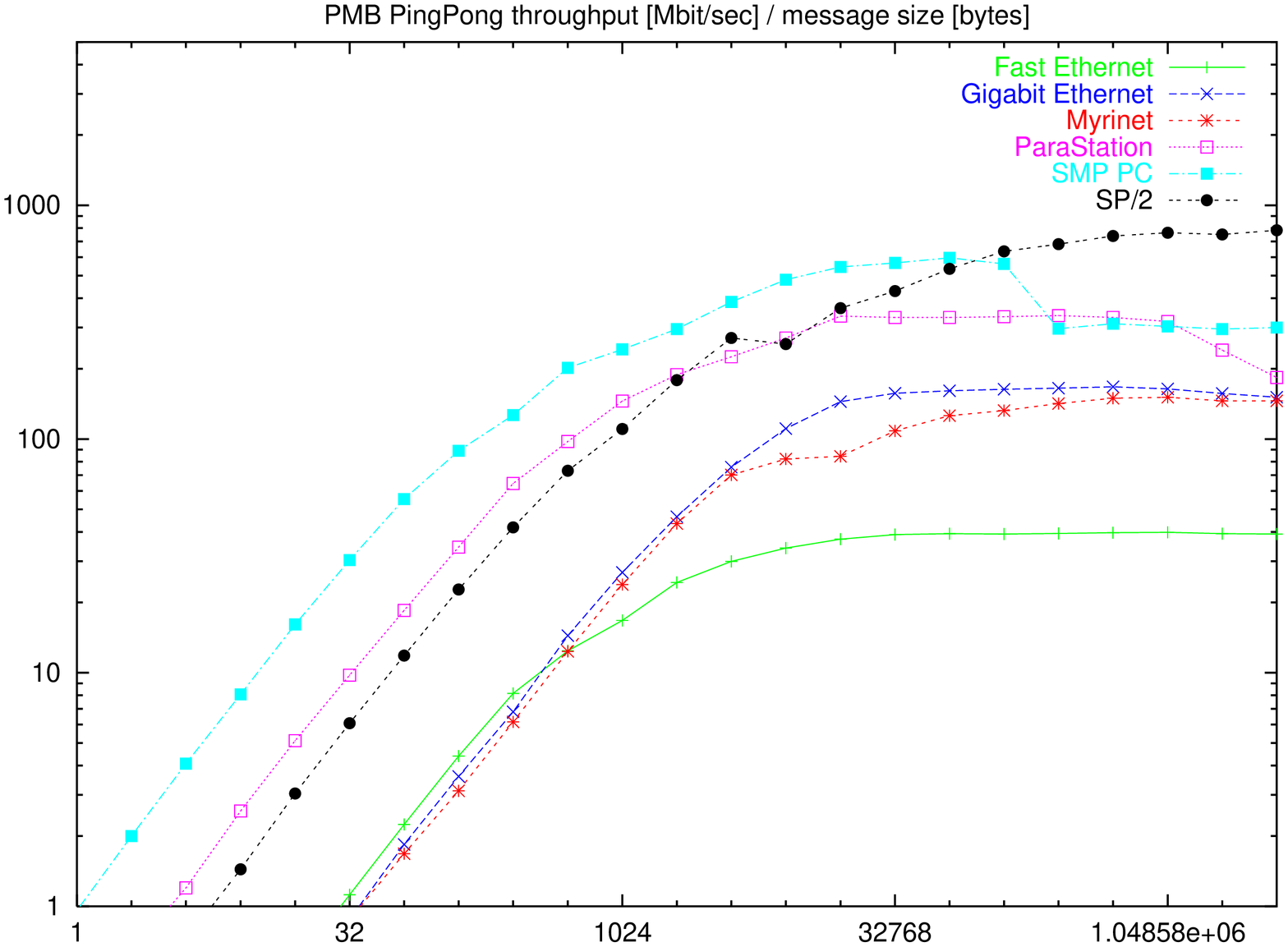,height=7.5cm,angle=-90}
\epsfig{file=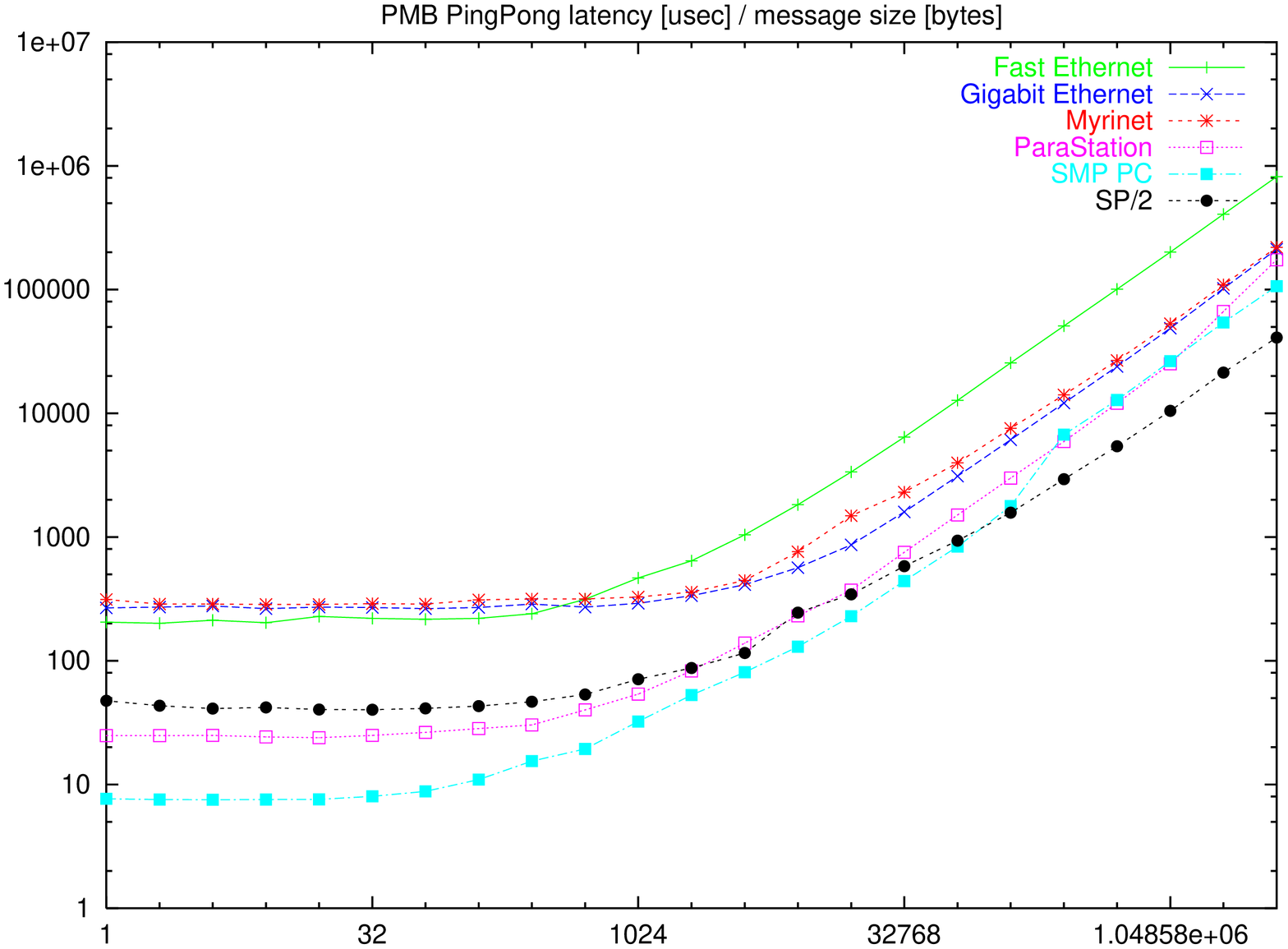,height=7.5cm,angle=-90}
\end{center}
\caption{MPI(CH) bandwidth and latency results for the PingPong benchmark.}
\label{pingpong}
\end{figure}

\noindent
As can be seen from figure \ref{pingpong} (left) the bandwidth increases as 
the message size does. For large message sizes there is a saturation effect. 
The maximum throughput for the Fast Ethernet is about 40 MBit/s, for the
Gigabit Ethernet and the Myrinet the maximum bandwidth is about 200 Mbit/s.
The ParaStation II software provides a special device driver for MPI(CH) on 
the Myrinet hardware that does not use the IP protocol and adds much less 
protocol overhead. With this spezialized software a bandwidth of about 
300 MBit/s is achieved. The IBM SP2 reaches a maximum MPI bandwidth of 800 
MBit/s, which is even higher than the maximum MPI(CH) throughput on the SMP
machine that uses shared memory segments for the data transfer.
\noindent
In figure \ref{pingpong} (right) the measured MPI(CH) latency for the 
ping-pong benchmark is shown. The latency ranges from a few microseconds for
small message sizes up to a second for large message sizes. Using MPI(CH)
over the IP protocol results in a minimum latency of more than 200 
microseconds.
Lower latencies can be achieved with special device drivers only. It turns 
out that the minimum latency on the ParaStation II (about 20 microseconds) is 
even smaller than on the IBM SP2. Of course the SMP machine has the smallest
latency---less than ten microseconds. Note that for our particular interest
the latency at larger message sizes is most important. In this region again
the IBM SP2 provides optimal performance.


\vspace{5mm}
\noindent
{\large\bf 3. The Parallelization of FORM}
\vspace{5mm}

\noindent
Allowing local operations only makes FORM very well suited for a
straightforward parallelization: distribute the input terms among the 
available processors, let each of them perform the local operations on its
input terms and generate and sort the arising output terms. At the end of 
a module the sorted streams of terms from all processors have to be merged
into one final output stream again. The compiling of the program-text to the
internal representation was considered not to be worth parallelizing. 
This concept indicates to use a master-slave structure for the 
parallelization, where the master would store the expressions and distribute
and recollect all the terms of each expression.
For the implementation of this raw concept we used a four step strategy:
\begin{itemize}
\item one process(or) generates terms, a second process(or) sorts the 
output terms
\item instead of only one process arbitrary many processes perform the 
sorting.
\item the input terms are distributed and the term generation is also done
in parallel
\item final optimizations: avoid or handle worst cases, load leveling,
fault tolerance
\end{itemize}
This approach has several advantages, the most important being that having 
working versions in every stage gives us a good idea of how good the 
parallelization is and the possibility of doing realistic tests even 
in a very early stage.


\vspace{5mm}
\noindent
{\large\bf 3.1 The Two-Processor Version}
\vspace{5mm}

\noindent
This first step turns out to be useful, since it not only gives a deep 
insight into how changes to the source code of FORM have to be made without
affecting the efficiency of the well optimized sequential code. 
The two-processor version also serves as a check of whether and how the 
concept can possibly lead to a decent speedup, because we basically add 
communication overhead (no speedup can be expected from seperating the
generation and the sorting of terms only). 
It turns out that for parallelizing software on a cluster of very fast 
workstations the importance of avoiding communication overhead cannot be 
overstated. Moreover these experiments show that the communication has 
to be done in a buffered way, since sending single terms increases the 
runtimes of the two-processor code up to a factor 20. With the buffered 
version the increase in runtime was limited to about a factor of 1.5 of
the sequential code with two workstation connected by a (slow) 10 Mbit/s
Ethernet using the PVM and MPI(CH) libraries.


\vspace{5mm}
\noindent
{\large\bf 3.2 Parallel Sorting}
\vspace{5mm}

\noindent
The second step is to distribute the output terms among arbitrary many 
processors and do the sorting in parallel. Since this part of the sorting 
relies strongly on communication between the processors, it most probably
sets the limits of parallel speedup. Therefore already in this stage it 
could be predicted quite reliably whether a speedup of the whole program
could be achieved or not. 
A first try was to map the ``tree of losers'' used in FORM to merge sorted 
patches onto the processors. While it would distribute the workload in an
optimal way this approach adds too much communication overhead. 
This is why in the end a much simpler communication structure was used,
where all slaves send their sorted terms to the master process and this 
process uses a local ``tree of losers'' to merge the output streams of the
slave-processes. 
Additional effort was made to overlap the work on the master process with the 
sorting done on the slaves, which caused a much deeper interference with the 
sequential code, but finally resulted in a very fast and stable
implementation.

Of course the scaling of this approach with a large number of processors is 
a possible limitation, but tests on the 256 node IBM SP2 parallel computer
showed that, at least for such a specialized network, the performance 
does not dramatically break down up to 32 processors. 
If problems should occur with more processors or a different architecture, 
an intermediate layer of ``foremen'' could be used to circumvent this 
possible bottleneck.

\begin{figure}[h!]
\begin{center}
\epsfig{file=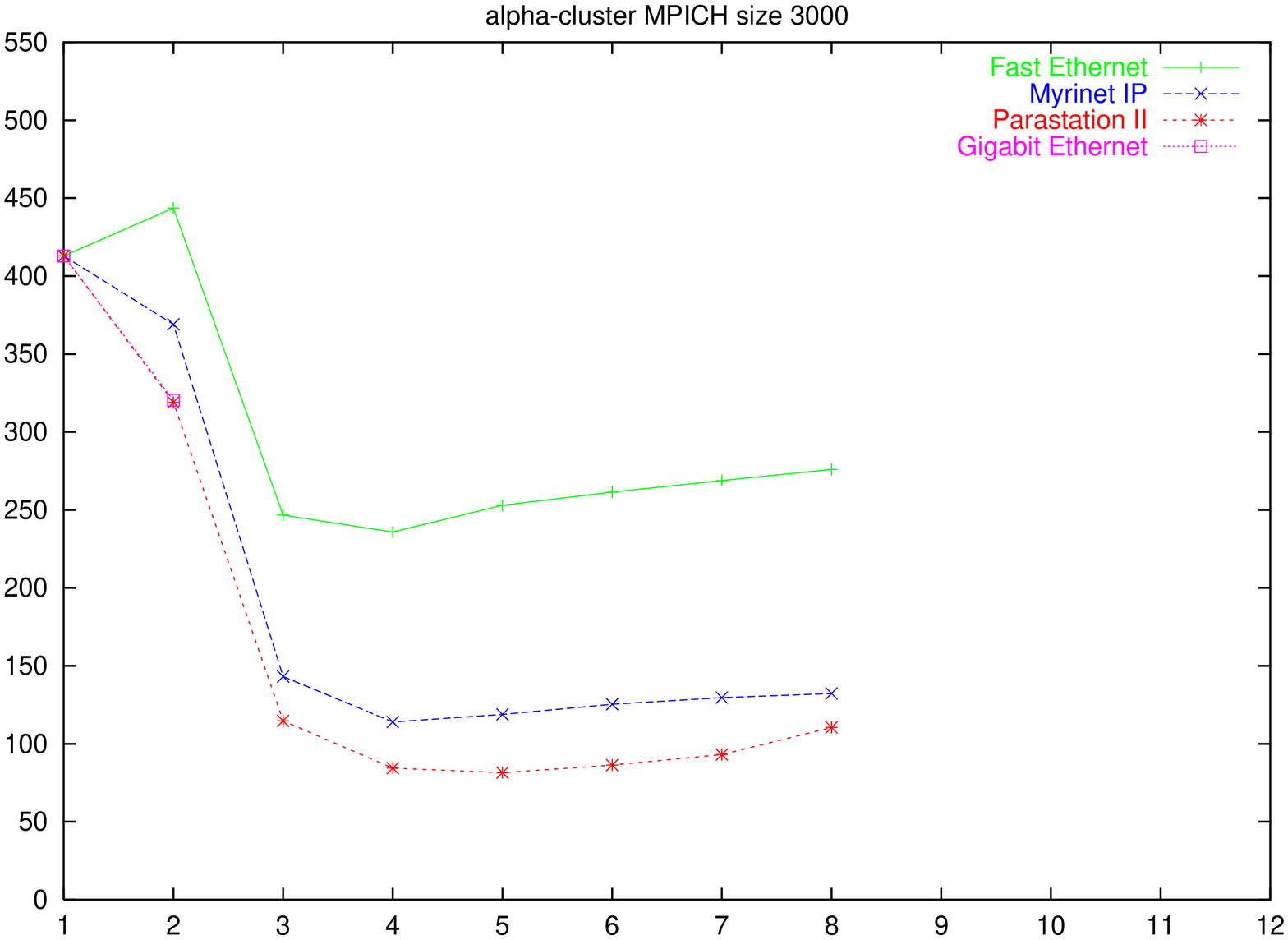,height=7.5cm,angle=-90}
\epsfig{file=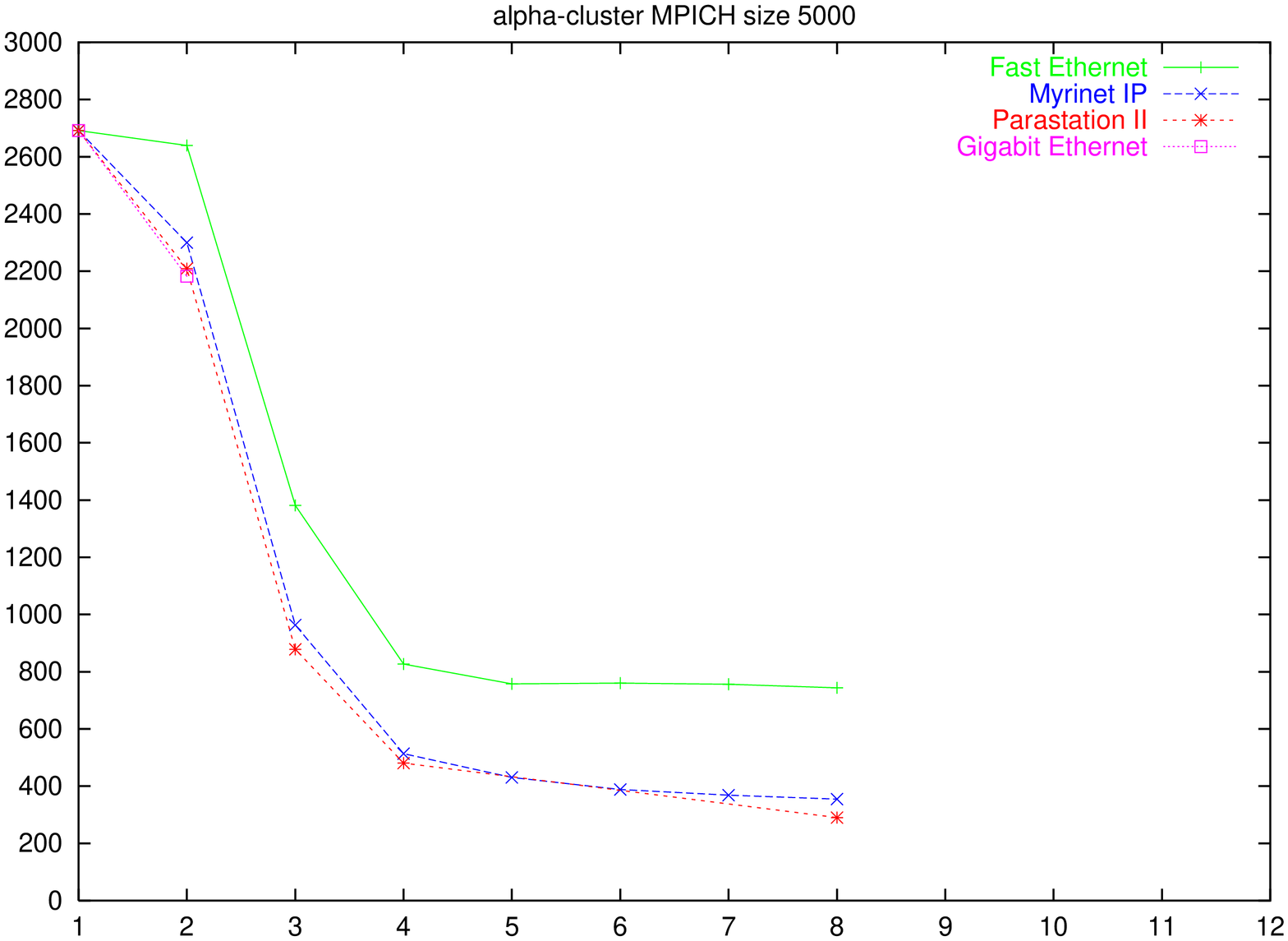,height=7.5cm,angle=-90}
\end{center}
\caption{Runtimes [sec] for parallel sorting on the alpha cluster,
shown is the runtime vs. number of processors for different 
communication soft/hardware.
}
\label{sortplots}
\end{figure}

Since we are only testing the sorting, any problem that produces a sufficient
number of intermediate terms can be used to test the parallel speedup. We 
wrote a very simple program, that expanded the expression 
$(a_1+\cdots+a_n)^2$ and then replaced $a_1$ by $-(a_4+\cdots+a_n)$ which
results in a short, easy to check result and, by choosing different values 
of $n$, can be scaled in an easy way. The runtimes we could achieve with this
version on the alpha cluster with different combinations of communication
soft/hardware are shown in figure \ref{sortplots} for 
a medium ($n=3000$, $\sim 500$ MB) 
and a large ($n=5000$, $\sim 1400$ MB) problem (the sizes in MB are for the 
64-bit architecture of the alphas, they are half as large for 32 bit systems
(a more efficient compression to reduce disk access is in progress). 
They show that only for sufficiently large problems a satisfying speedup can
be achieved without sophisticated network soft/hardware.

\begin{figure}[h!]
\begin{center}
\epsfig{file=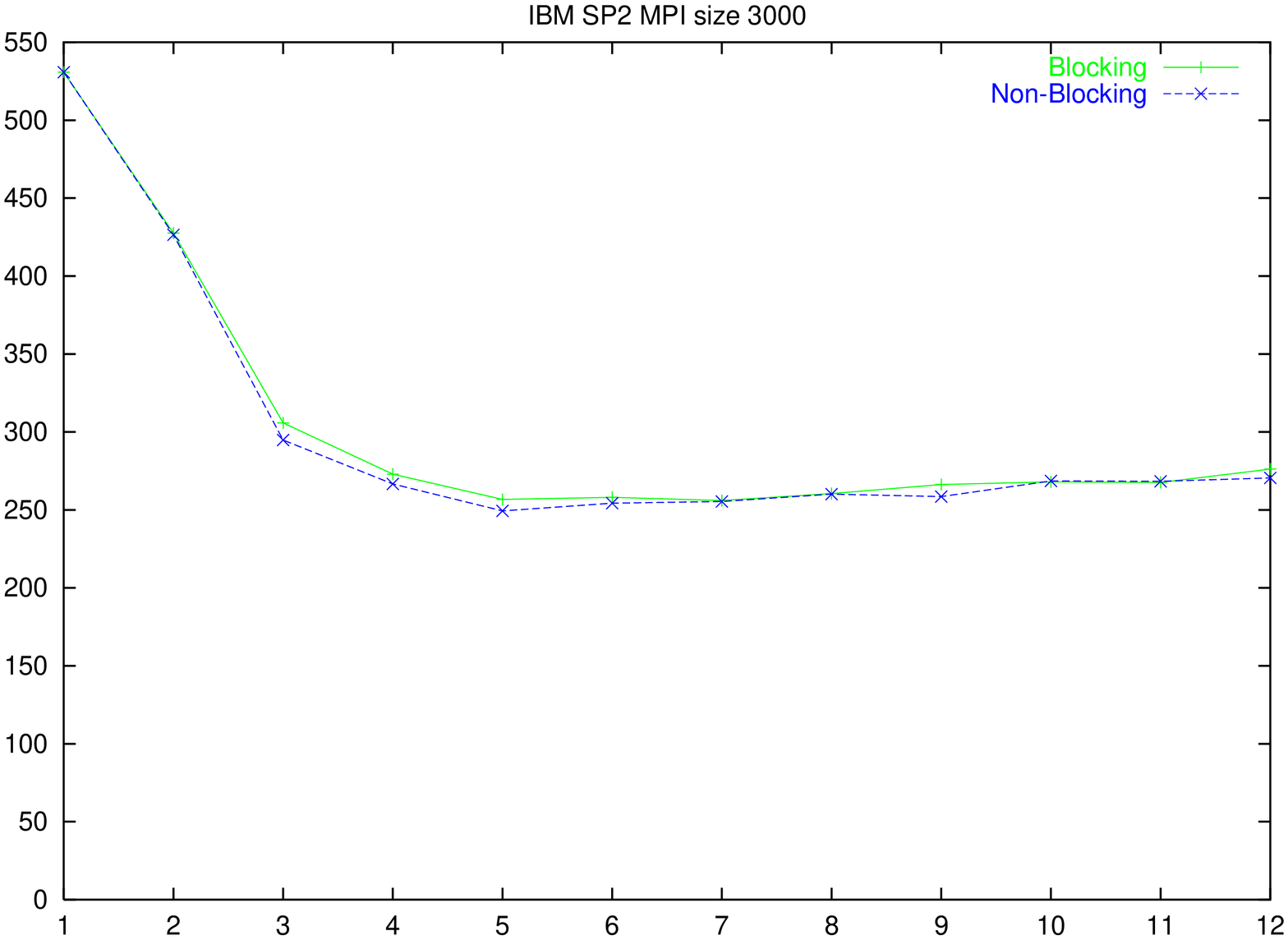,height=7cm,angle=-90}
\epsfig{file=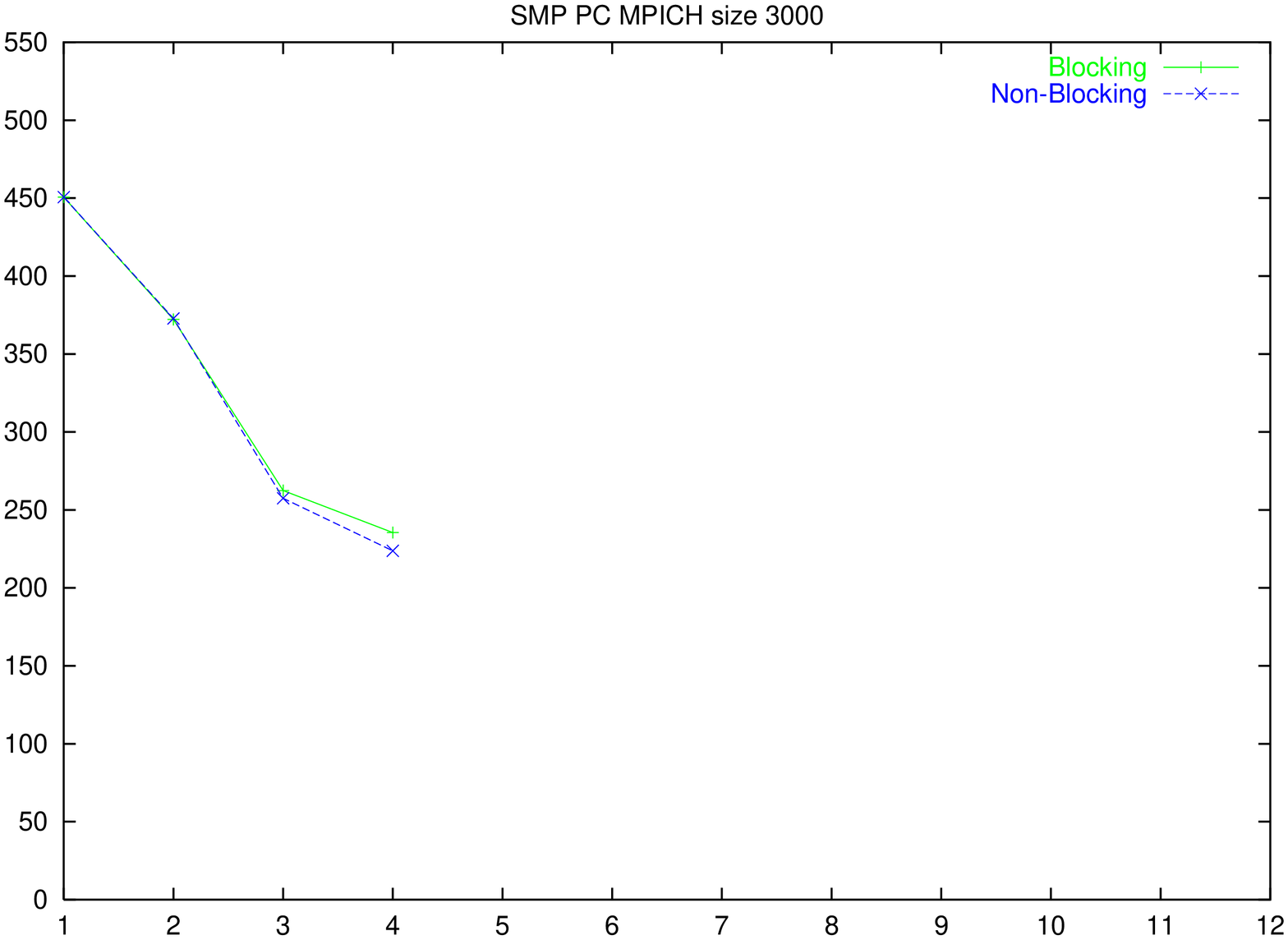,height=7cm,angle=-90}\\
\end{center}
\caption{Runtimes [sec] for parallel sorting on two other architectures:
IBM SP2 (left)
and Intel SMP (right)
shown are the
version with blocking 
and nonblocking 
communication.
}
\label{sp2-shm}
\end{figure}

\noindent
The parallel sorting was implemented in a way that can take advantage of 
the non-blocking unbuffered MPI message passing functions by using a set
of cyclic buffers which eliminates any unnecessary overhead and also 
minimizes the time consumed for synchronizing. 
Choosing the number of cyclic buffers to be one results in using the 
send/receive function in blocking mode. 
On a 4 processor SMP machine we could see a speedup of about 10\% for some 
cases using this feature, while on a IBM SP2 there are no differences between 
the two versions (figure \ref{sp2-shm}).

\noindent
It is also quite interesting to see that obviously the master process can 
generate just enough terms to keep about 4 processes busy, almost 
independently of the problem size, the difference is that the cost of 
additional overhead for more than 4 processes is larger for
the small problems, which of course is no surprise.
It was know from the profiling of the sequential code that a large part of 
the runtime is spent in the generation routines, so the speedup we could see
for very large problems is about the optimum we could expect to reach at this 
intermediate stage.

\vspace{5mm}
\noindent
{\large\bf 3.3 Parallel Generating}
\vspace{5mm}

\noindent
The last step towards a working prototype was the distribution of input 
terms among all processes before generation of the output terms. 
The slave-processes of course also need all the information necessary 
for the generation of the terms, which is at the moment realized by having 
them all read the program text file and compile their own internal 
representation and broadcasting only the rest of necessary information 
from the master process. 

It was one of the main goals to get this version to run realistic problems
as soon as possible, so the often used FORM-package MINCER \cite{mincer},
which can calculate certain types of Feynman diagrams up to three loops was 
used to serve as a testbed.
First the necessary subset of FORM commands used in this package was made 
available and some easy standard integrals served as a test of the correctness
of the parallel computations. After these preliminary tests the computation of
some diagrams of an ongoing project, the calculation of the 3-loop 
triple-gluon vertexfunction, are now serving as ``real world'' tests.

\begin{figure}[h!]
\begin{center}
\epsfig{file=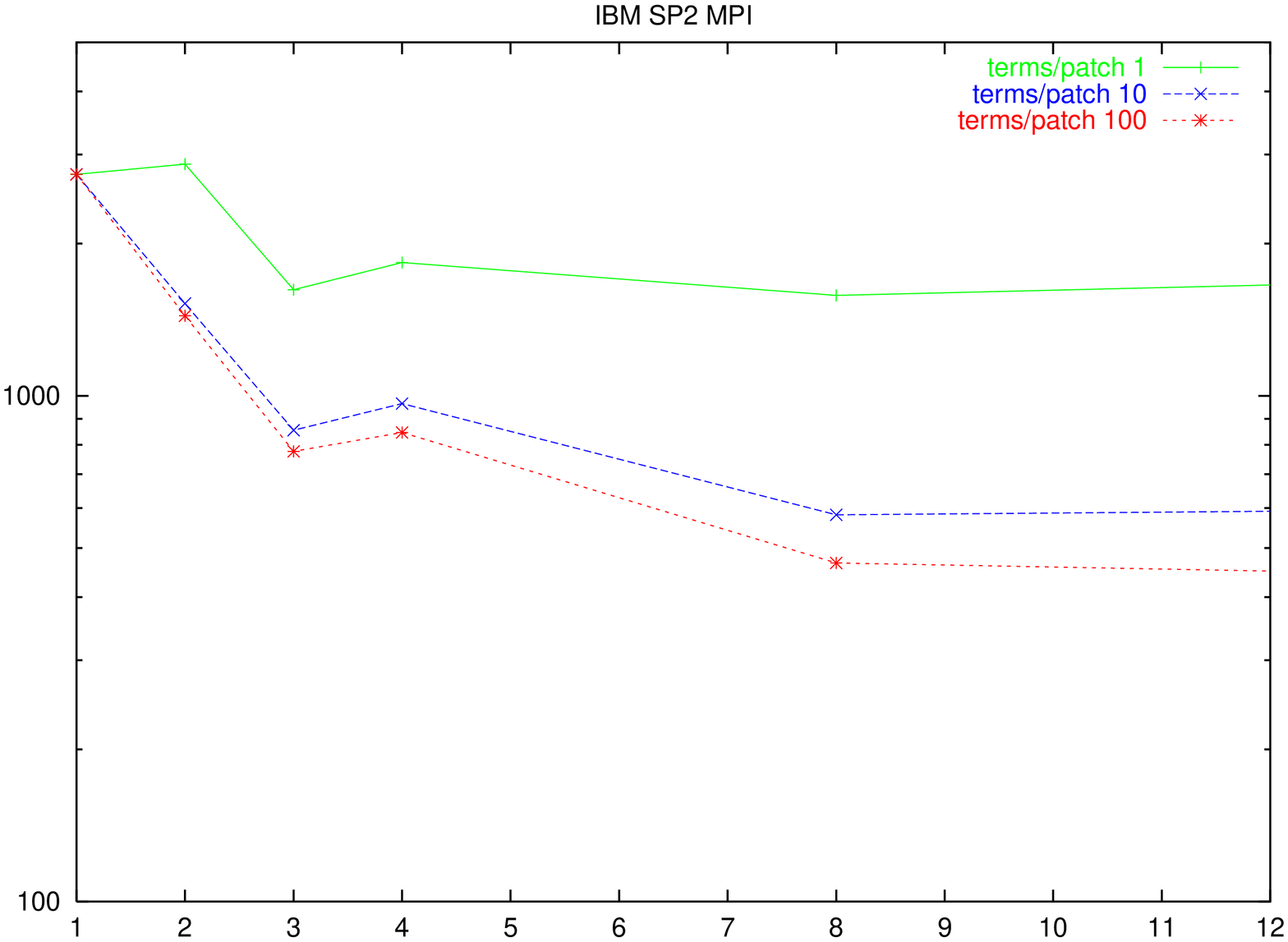,height=7cm,angle=-90}
\epsfig{file=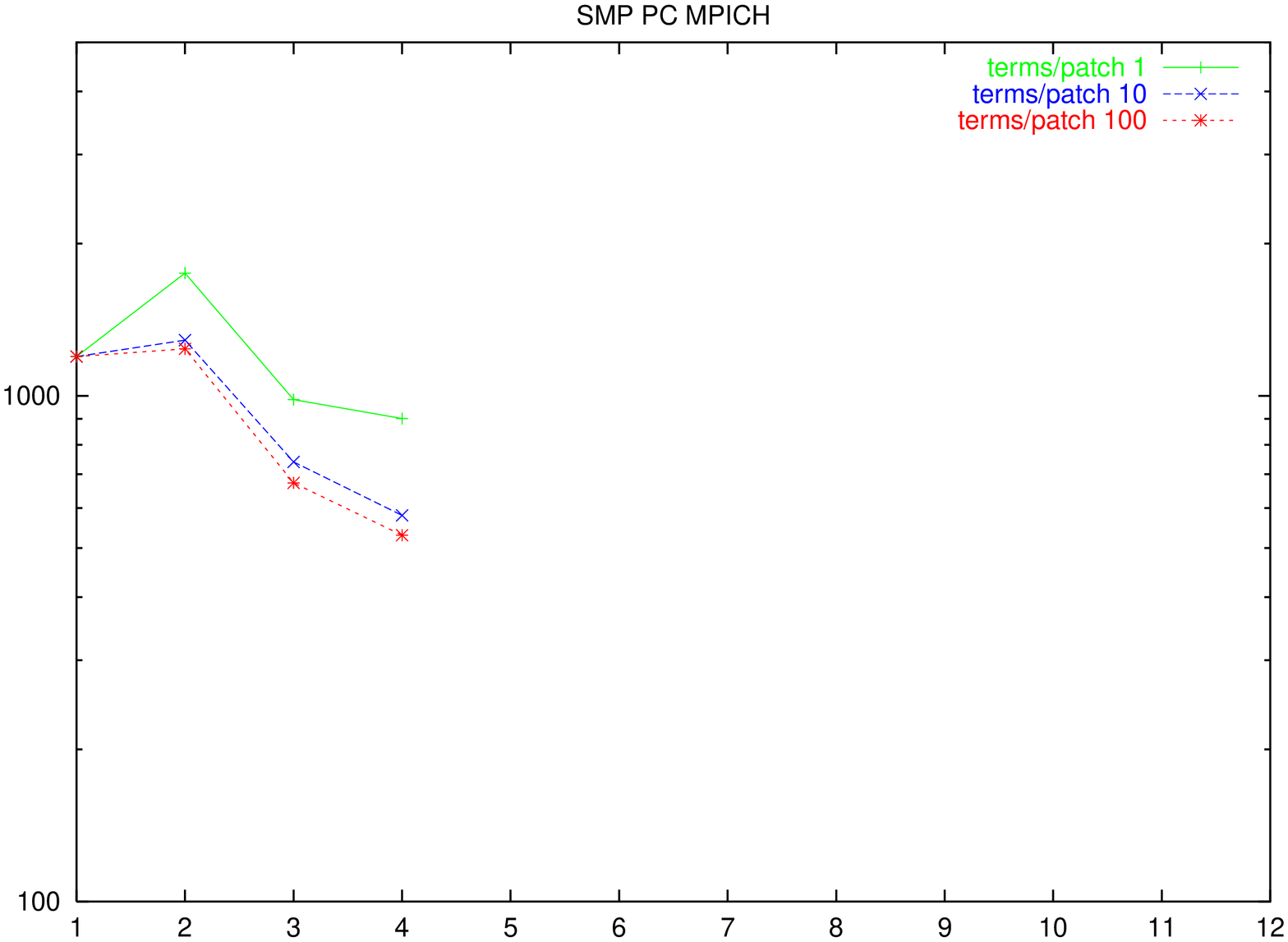,height=7cm,angle=-90}
\end{center}
\caption{Runtimes [sec] vs.~number of processors for a ``simple'' 3-loop 
Feynman-diagram on a IBM SP2 (left) and a 4 processor SMP machine (right)
using its proprietary MPI-library. Shown are runtimes for different 
granularities of the input-term distribution.}
\label{pargensp2smp}
\end{figure}

Just as with the sorting the speedup is strongly dependent on the number of 
terms sent within one message. 
In the current implementation this number can be changed at the start of the 
program and results in very different speedups, as can be seen in figure 
\ref{pargensp2smp}. 
Of course choosing a too coarse grained distribution will result in the 
danger of running into worst cases, where all the work sits in only one of 
the input patches, and only one processor is busy.
On the other hand, the fine grain distribution causes more overhead. 
The best setting turned out to be not only dependent on the underlying 
soft/hardware, but also strongly dependent on the problem that is run. 
The distribution is organized such that the master sends a patch of terms
to each slave process at the beginning of the module and then waits for the 
slaves to ask for new terms whenever they are finished with their last patch 
of input terms. This actually turns the concept in that of a client-server
situation which will also be useful to make the slaves receive any kind 
of global information when the need arises. 
It also produces a decent load levelling among the slaves, which can be 
controlled by the size of the input-patches and could even be adjusted
during runtime for further improvement.
 
It must be understood that for this specific problem there were now over 
hundred modules executed, where most of them were only seeing 1 to 10 terms, 
which makes them ``worst cases'' for the parallelization and there is a lot
of room for further optimizations of this parallel program. 
Also the results are received from the FORM-code written for the 
sequential version of FORM without any modification or optimization in these
algebraic packages, which corresponds to a perfect code reuse. 

\noindent
As was expected, on the slow nodes of the IBM SP2 with fast connections a
speedup is quite easy to obtain (see figure \ref{pargensp2smp}, left). 
Also it can be seen that for the fully parallelized program there is indeed
a speedup up to 12 processors on the SP2. Another result is that the number 
of input terms that are distributed at once has a large influence on the 
speedup that can be achieved. 
Figure \ref{pargensp2smp} (right) shows the runtimes that could be achieved 
on a 4 processor SMP machine. Also on that architecture the number of input 
terms sent at a time must be at least about 10 to see a speedup. 

\begin{figure}[h!]
\begin{center}
\epsfig{file=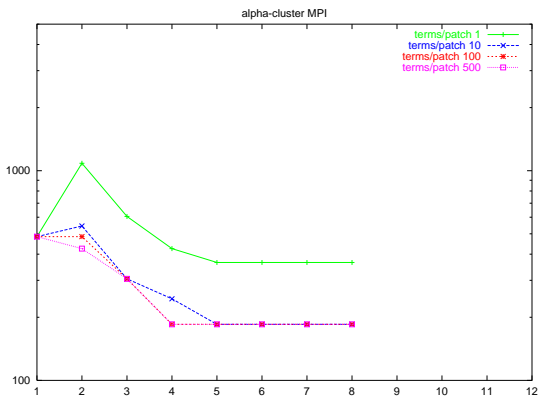,height=7.5cm,angle=-90}
\end{center}
\caption{Runtimes [sec] vs.~number of processors for a ``simple'' 3-loop 
Feynman-diagram on the DEC alpha cluster using MPI(CH) over the 
ParaStation II hard/software. Shown are runtimes for different 
granularities of the input-term distribution.}
\label{pargenalpha}
\end{figure}

\noindent 
As can be seen in figure \ref{pargenalpha} this holds also for the DEC alpha
cluster using MPI(CH) over the Para\-Station II hard/software. Obviously a 
speedup is much harder to obtain in this case. Still the parallel version on
the IBM SP2 and the 4-processor SMP machine does hardly reach the runtimes 
measured with the sequential version on the DEC alpha cluster.
From our experience with the parallel sorting and preliminary tests we 
expect a much larger speedup for larger problems, which are of course 
most interesting and investigated at the moment. 


\vspace{5mm}
\noindent
{\large\bf 4. Conclusion \& Outlook}
\vspace{5mm}

\noindent
The ongoing work is mainly to find and fix bugs to make the parallel version
to run larger problems just as extremely reliable as the sequential one, 
which has been tested and improved for over ten years. 
As the runtimes show, there is also some work to be done to further optimize
the parallel version, especially to insure that modules with only few terms
will not cause a slowdown by unnecessary communication overhead.
The next step will be the implementation of the full FORM version 2.3
standard, so that all existing software for that version can be run in 
parallel (which is basically all software that exists for FORM). Since the 
parallel program is actually based on version 3.0 which is in preparation by 
the author J. Vermaseren and offers some new and powerful features, we also 
will investigate if and how these new features can be implemented in the 
parallel version. 
Another field of current and planned activities is porting the program to 
other architectures. We are especially interested in taking better 
advantage of the possibilities of SMP machines.
The main goal of all these efforts is to get from the current stage of a 
working prototype to an easy to use, powerful and reliable program that
is not an end in itself, but a useful tool in real life applications on a 
wide variety of (parallel) architectures in the same manner as the current 
sequential version of FORM.

This work has been supported by DFG Forschergruppe under contract 
no.~KU 502/8-2 and Digital Equipment under contract no.~DE-98008.




\begin{thebibliography}{99}
\bibitem{form}{J.A.M. Vermaseren, {\it Symbolic Manipulation
with FORM}, published by CAN (Computer Algebra Nederland), Kruislaan
413, 1098 SJ Amsterdam, 1991, ISBN 90-74116-01-9}
\bibitem{knuth}{D.Knuth, \textit{The Art of Computer Programming},
Addison-Wesley, 1997.}
\bibitem{mpich} MPI(CH) homepage: http://www.mcs.anl.gov/mpi/
\bibitem{pvm} PVM homepage: http://www.epm.ornl.gov/pvm/
\bibitem{myricom} Myricom homepage: http://www.myri.com.
\bibitem{parastation2} ParaStation homepage: 
http://wwwipd.ira.uka.de/ParaStation/PSM/
\bibitem{netperf}{Netperf homepage: 
http://www.netperf.org/nerperf/NetperfPage.html}
\bibitem{pmb}{Pallas MPI Benchmarks: http://www.pallas.de/PMB2/}
\bibitem{mincer}{S.A.Larin, F.V.Tkachov, J.A.M. Vermaseren, Preprint
NIKHEF-H/91-18 (1991)}
\end{thebibliography}
\end{document}